\documentclass[useAMS,usenatbib]{mn2e}

\usepackage{xspace}
\usepackage{aas_macros}
\usepackage{graphicx}
\usepackage{color,url}
\usepackage{amsmath}

\def\deg{\ensuremath{^\circ}\xspace}

\def\kms{\ensuremath{{\rm km\, s}^{-1}}\xspace}
\def\SiIII{Si\,{\sc iii}\xspace}
\def\SiII{Si\,{\sc ii}\xspace}
\def\HeI{He\,{\sc i}\xspace}
\def\NeII{Ne\,{\sc ii}\xspace}

\title[Multisite spectroscopic seismic study of the $\beta\,$Cep star V2052~Oph: inhibition of mixing by its magnetic field]{Multisite spectroscopic seismic study of the $\beta\,$Cep star V2052~Ophiuchi: inhibition of mixing by its magnetic field}

\author[M. Briquet et al.]
{\parbox{\textwidth}{M. Briquet,$^{1,2,3}$\thanks{E-mail: maryline.briquet@ulg.ac.be}\thanks{F.R.S.-FNRS Postdoctoral Researcher, Belgium} 
C.~Neiner,${^3}$
C.~Aerts,$^{2,4}$
T.~Morel,${^1}$
S.~Mathis,${^{5,3}}$
D.R.~Reese,${^1}$ 
H.~Lehmann,${^6}$ 
R.~Costero,${^7}$ 
J.~Echevarria,${^7}$ 
G.~Handler,${^8}$ 
E.~Kambe,${^9}$ 
R.~Hirata,${^{10}}$
S.~Masuda,${^{11}}$
D.~Wright,${^{12}}$
S.~Yang,${^{13}}$\thanks{Guest investigator, Dominion Astrophysical Observatory, Herzberg Institute of Astrophysics, National Research Council of Canada},
O.~Pintado,${^{14}}$ 
D.~Mkrtichian,${^{15,16}}$ 
\mbox{B.-C.}~Lee,${^{17}}$ 
I.~Han,${^{17}}$ 
A.~Bruch,${^{18}}$ 
P.~De Cat,${^{19}}$
K.~Uytterhoeven,${^{20,21}}$
K.~Lefever,${^{2,22}}$ 
J.~Vanautgaerden,${^2}$
B.~de Batz,${^3}$
Y.~Fr\'emat,${^{19}}$
H.~Henrichs,${^{23}}$ 
V.C.~Geers,${^{24}}$
C.~Martayan,${^{25}}$
A.M.~Hubert,${^{26}}$
O.~Thizy${^{27}}$ and
A.~Tijani${^{28}}$}\\
\\
\parbox{\textwidth}{$^1$Institut d'Astrophysique et de G\'eophysique Universit\'e de Li\`ege, All\'ee du 6 Ao\^ut 17, B-4000 Li\`ege, Belgium\\
$^2$Institute of Astronomy - KULeuven, Celestijnenlaan 200D, 3001 Leuven, Belgium\\
$^3$LESIA, UMR 8109 du CNRS, Observatoire de Paris, UPMC, Univ. Paris Diderot, 5 place Jules Janssen, 92195 Meudon Cedex, France \\
$^{4}$Department of Astrophysics, University of Nijmegen, PO Box 9010, 6500 GL Nijmegen, The Netherlands\\
$^5$Laboratoire AIM Paris-Saclay, CEA/DSM-CNRS-Universit\'e Paris Diderot; IRFU/SAp, Centre de Saclay, 91191 Gif-sur-Yvette Cedex, France \\
$^6$Th\"uringer Landessternwarte Tautenburg (TLS), Sternwarte 5, 07778 Tautenburg, Germany\\
$^7$Instituto de Astronom\'ia, Universidad Nacional Aut\'onoma de M\'exico, Ciudad Universitaria, M\'exico, D.F., 04510, M\'exico\\
$^8$Copernicus Astronomical Center, Bartycka 18, 00-716 Warsaw, Poland\\ 
$^9$Okayama Astrophysical Observatory, National Astronomical Observatory of Japan, Honjyo 3037-5, Kamogata, Asakuchi, Okayama, 719-0232, Japan\\
$^{10}$Otokoyama-Yutoku 7, E17-201, Yawata, Kyoto, 614-8371, Japan\\ 
$^{11}$Hayashi 2528-5, Takamatsu, Kagawa, 761-0301, Japan\\ 
$^{12}$Department of Astrophysics, School of Physics, University of New South Wales, NSW 2050, Australia\\ 
$^{13}$Department of Physics and Astronomy, University of Victoria, PO Box 3055, Victoria BC V8W 3P6, Canada\\
$^{14}$Instituto Superior de Correlaci\'on Geologica-CONICET, Tucuman, Argentina\\
$^{15}$National Astronomical Research Institute of Thailand,  Siripanich Building 191 Huay Kaew Road, Muang District, Chiangmai, Thailand 50200\\
$^{16}$Crimean Astrophysical Observatory, Nauchny, Crimea, 98409, Ukraine\\
$^{17}$Korea Astronomy and Space Science Institute, 776, Daedeokdae-Ro, Youseong-Gu, Daejeon 305-348, Korea\\
$^{18}$Laborat\'orio Nacional de Astrof\'isica, Rua Estados Unidos, 154 -- Bairro das Nac\~oes, 37504-364 Itajub\'a, Brazil\\
$^{19}$ Royal Observatory of Belgium, 3 avenue circulaire, 1180 Brussel, Belgium\\
$^{20}$Instituto de Astrof\'{\i}sica de Canarias (IAC), Calle Via Lactea s/n, 38205 La Laguna, Tenerife \\
$^{21}$Dept. Astrof\'{\i}sica, Universidad de La Laguna (ULL), Tenerife \\
$^{22}$Belgian Institute for Space Aeronomy, Ringlaan 3, 1080 Brussels, Belgium\\
$^{23}$Astronomical Institute Anton Pannekoek, University of Amsterdam, Science Park 904, 1098XH Amsterdam, Netherlands\\
$^{24}$ETH Zurich, Institute for Astronomy, Wolfgang-Pauli-Strasse 27, 8093 Zurich, Switzerland\\
$^{25}$European Organisation for Astronomical Research in the Southern Hemisphere, Alonso de Cordova 3107, Vitacura, Santiago de Chile, Chile\\ 
$^{26}$GEPI, UMR 8111 du CNRS, Observatoire de Paris, Univ. Paris Diderot, 5 place Jules Janssen, 92195 Meudon Cedex, France \\
$^{27}$Shelyak Instruments, Les Roussets, 38420 Revel, France\\
$^{28}$Astronomical Institute ``Anton Pannekoek'', University of Amsterdam, Kruislaan 403, 1098 SJ Amsterdam, the Netherlands
\vspace{50mm}
}}

\begin{document}

\date{Accepted 2012 August 15. Received 2012 August 3; in original form 2012 June 15}

\pagerange{\pageref{firstpage}--\pageref{lastpage}} \pubyear{2012}

\maketitle

\label{firstpage}


\begin{abstract}
We used extensive ground-based multisite and archival spectroscopy to derive observational constraints for a seismic modelling of the magnetic $\beta$~Cep star V2052~Ophiuchi. The line-profile variability is dominated by a radial mode ($f_1$=$7.14846$ d$^{-1}$) and by rotational modulation ($P_{\rm rot}$=$3.638833$ d). Two non-radial low-amplitude modes ($f_2$=$7.75603\ \rm{d}^{-1}$ and $f_3$=$6.82308\ \rm{d}^{-1}$) are also detected. The four periodicities that we found are the same as the ones discovered from a companion multisite photometric campaign (Handler et al.\,\citealp{handler12}) and known in the literature. Using the photometric constraints on the degrees $\ell$ of the pulsation modes, we show that both $f_2$ and $f_3$ are prograde modes with $(\ell,m)$=$(4,2)$ or $(4,3)$. These results allowed us to deduce ranges for the mass ($M \in [8.2,9.6]\ $M$_\odot$) and central hydrogen abundance ($X_c \in [0.25,0.32]$) of V2052~Oph, to identify the radial orders $n_1$=$1$, $n_2$=$-3$ and $n_3$=$-2$, and to derive an equatorial rotation velocity $v_{\rm eq} \in [71,75]$ km~s$^{-1}$. The model parameters are in full agreement with the effective temperature and surface gravity deduced from spectroscopy. Only models with no or mild core overshooting ($\alpha_{\rm ov} \in [0,0.15]$ local pressure scale heights) can account for the observed properties. Such a low overshooting is opposite to our previous modelling results for the non-magnetic $\beta$~Cep star $\theta$~Oph having very similar parameters, except for a slower surface rotation rate. We discuss whether this result can be explained by the presence of a magnetic field in V2052~Oph that inhibits mixing in its interior. 
\end{abstract}


\section{Introduction}
Stellar modelling of B-type main-sequence stars is benefiting greatly from seismic data of $\beta\,$Cep stars. Among the various classes of B-type pulsators (e.g. Chapter\,2 of Aerts et al.\,\citealp{aerts10}) this is the only one with members having clearly identified values for the wavenumbers $(\ell,m,n)$ for several of the detected oscillation modes, which is a prerequisite for successful seismic modelling. 


Models based on standard input physics could satisfactorily explain seismic data of the $\beta\,$Cep stars V836\,Cen (Aerts et al.\,\citealp{aerts03}; Dupret et al.\,\citealp{dupret04}), $\beta\,$CMa (Mazumdar et al.\,\citealp{mazumdar06}), $\delta\,$Ceti (Aerts et al.\,\citealp{aerts06}), and $\theta\,$Oph (Briquet et al.\,\citealp{briquet07}).
However, models with the same input physics failed to explain the much richer seismic constraints assembled from huge multi-technique multisite campaigns organised during many months for the $\beta\,$Cep stars $\nu\,$Eri (De Ridder et al.\,\citealp{deridder04}; Pamyatnykh, Handler \& Dziembowski\,\citealp{pamyatnykh04}; Ausseloos et al.\,\citealp{ausseloos04}) and 12\,Lac (Handler et al.\,\citealp{handler06}; Dziembowski \& Pamyatnykh\,\citealp{dziembowski08}; Desmet et al.\,\citealp{desmet09}). These campaigns led to more than twice the number of modes known previously in those two stars, among which there were one or two high-order $g$-modes, and revealed shortcomings in the excitation predictions for some modes. 

A similar excitation problem occurred for $\gamma\,$Peg, a class member found to pulsate in several high-order $g$-modes and low-order $p$-modes from MOST space-based photometry (Handler et al.\,\citealp{handler09}; Walczak \& Daszy\'nska-Daszkiewicz\,\citealp{walczak10}). Zdravkov \& Pamyatnykh\,\citet{zdravkov09} suggested that an increase in the opacities by 20\% might solve the excitation problem of $\gamma\,$Peg. Another excitation problem was pointed out for one mode of the CoRoT target V1449 Aql (Aerts et al.\,\citealp{aerts11}). The recent detection and interpretation of $\beta\,$Cep-type modes in the O9V pulsator HD\,46202 (Briquet et al.\,\citealp{briquet11}) with CoRoT observations showed that the excitation problem of massive pulsators is acute, since none of the detected modes is predicted to be excited for appropriate stellar models representing this star. 
In all these studies, the effects of rotation on the excitation and amplitude of modes were not or only partly treated. Non-adiabatic computations taking rotation into account of the excitation of modes are available (e.g. Lee\,\citealp{lee98}; Townsend\,\citealp{townsend05}) but those of the amplitude of modes are just starting to become available (Lee\,\citealp{lee12}). These calculations will be very useful to check whether a better agreement between observed and theoretically excited modes may be found. 

An important result of these asteroseismic studies is the evidence for non-rigid interior rotation in some $\beta$~Cep stars. It was shown that V836~Cen, $\nu$~Eri and 12~Lac rotate more rapidly in their inner parts than at their surface. These works also allow us to test whether core overshooting has to be included in our stellar models. Overshooting represents here the amount of non-standard mixing processes (Maeder\,\citealp{maeder09}; Mathis\,\citealp{mathis10}). For most of the $\beta$~Cep targets modelled so far, the conclusion was the need for core overshooting for a better agreement with the pulsational characteristics. The derived core overshooting parameter values are however small (around 0.10 local pressure scale heights for V836~Cen and HD\,46202, and around 0.20 for $\delta$~Ceti, $\beta$~CMa, and 12~Lac)
%
with the exception of $\theta$~Oph for which the value was found to be around 0.40. For two cases, $\nu$~Eri and V1449 Aql, the core overshooting parameter could be kept to zero.


Space-based observations revealed other pulsators than $\gamma\,$Peg with both $g$- and $p$-mode pulsations of SPB and $\beta$~Cep types (Balona et al.\,\citealp{balona11}). For the CoRoT hybrid HD~50230, the observation of deviations from a uniform period spacing led to the first exploration of the regions adjacent to the core in a massive star (Degroote et al.\,\citealp{degroote10}). HD~43317 was discovered by CoRoT to be another hybrid star with a wealth of pulsational constraints but also showing the presence of chemical inhomogeneities at its stellar surface (P\'apics et al.\,\citealp{papics12}). Before space-based asteroseismology, the simultaneous presence of both pulsation and rotational modulation was found from ground-based spectroscopy only in $\beta$~Cep (Telting, Aerts \& Mathias\,\citealp{telting97}), V2052~Oph (Neiner et al.\,\citealp{neiner03}), and $\kappa$~Sco (Uytterhoeven et al.\,\citealp{uytterhoeven05}). Another important discovery by means of space-based CoRoT data is the first observation of stochastically excited gravito-inertial modes in a massive star. Such modes have recently been detected in the hot (O9.5-B0) pulsator HD\,51452 (Neiner et al., submitted), and this type of excitation should probably be more widely considered for massive pulsators.

Among $\beta$~Cep stars with an asteroseismic modelling, a magnetic field has been detected so far in $\beta$~Cep (Henrichs et al.\,\citealp{henrichs00}; Shibahashi \& Aerts\,\citealp{shibahashi00}) and in V1449 Aql (Hubrig et al.\,\citealp{hubrig11}; Aerts et al.\,\citealp{aerts11}), although the field of the latter is still a matter of debate (Shultz et al.\,\citealp{shultz12}).
In order to understand the effect of a magnetic field on the seismic behaviour of $\beta\,$Cep stars,
a multisite photometric and spectroscopic campaign was set up for the equatorial star V2052~Oph ($m_V$=$5.8$, spectral type B2IV/V), which is known to be a magnetic pulsator with a dominant radial mode of frequency 7.145 d$^{-1}$ and a rotation frequency of 0.275 d$^{-1}$ (Neiner et al.\,\citealp{neiner03}). The star is slightly enriched in He, revealed a mild N excess (Morel et al.\,\citealp{morel06}), and was considered to be suitable as a seismic target, in view of the identified dominant radial mode 
and the discovery of an additional low-amplitude non-radial mode by Neiner et al.\,\citet{neiner03}.

The multisite photometry of V2052~Oph is reported in a companion paper by Handler et al.\,\citet{handler12}. Here, we present the spectroscopic part of the campaign. After a description of our dataset in Sect.\,\ref{sect_data}, we discuss the atmospheric parameters and chemical composition of the star (Sect.\,\ref{sect_parameters}). Sect.\,\ref{sect_lpv} is devoted to our frequency analysis and mode identification. Besides our line-profile study, we make a comparison with stellar models (Sect.\,\ref{sect_modelling}). We end with a discussion about the inhibition of mixing by magnetism and conclusions in Sect.\,\ref{sect_discussion} and Sect.\,\ref{sect_conclusions}, respectively.
  
\section{The data}\label{sect_data}

\begin{table*}
 \centering
 \begin{minipage}{180mm}
  \caption{Log of the spectroscopic data used in this paper. The Julian Dates are given in days, $\Delta T$ denotes the time-span expressed in days, $N$ is the
number of spectra and S/N denotes the average signal-to-noise ratio for each
observatory measured around the \SiIII line at 4552 \AA.}
\label{TAB:observatories}
  \begin{tabular}{@{}lrrccccccl@{}}
  \hline
  \hline
  Observatory  & Telescope & \multicolumn{2}{c}{Julian Date}  
& \multicolumn{3}{c}{Data amount and} &  Observer(s) \\
(Name of the instrument;&& Begin & End &    \multicolumn{3}{c}{quality}  & \\
resolution; wavelength range in \AA) && & & $\Delta T$  &  N & S/N & \\

 \hline
 \hline

2000-2004   &&\multicolumn{2}{c}{$-$2450000}&&&&\\
\hline
Pic du Midi, France  & 2.0-m TBL & 1730 & 3186 & 1457 & 161 & 165 & CN, HH, VG, AH, AT\\
\hspace{2mm}(MUSICOS; 35\,000; 4489--6619) &  &  &  & & & & &  &  \\
\hline
  2004   &&\multicolumn{2}{c}{$-$2450000} &&&&\\
\hline
 Pico dos Dias Observatory, Brazil 
&1.6-m& 3200 &3201 & 2 & 20 & 290& AB\\
\hspace{2mm}(Esp Coud\'e; 30\,000; 4479--4644)&&&&&&&&&\\
 Complejo Astron\'omico El Leoncito Observatory, Argentina 
&2.1-m&  3219& 3221 & 3 & 20 & 103 & OP\\
\hspace{2mm}(EBASIM; 40\,000; 3826--5759)&&&&&&&&&\\
 Bohyunsan Astronomical Observatory, Korea  
&1.8-m&  3133& 3161& 29& 37 & 186& DM, BL\\
\hspace{2mm}(BOES; 50\,000; 3751--9803)&&&&&&&&&\\
 Dominion Astrophysical Observatory, Canada 
&1.2-m& 3134& 3197& 64& 53 & 213& SY\\
\hspace{2mm}(45\,000; 4457--4603) && & & &&&\\
 McDonald Observatory, USA  & 2.7-m & 3195& 3198& 4 & 105 & 215& GH, ME\\
\hspace{2mm}(Coud\'e; 60\,000; 3619--10274) &  & & & & & & \\
 Okayama Astrophysical Observatory, Japan &
1.88-m & 3202& 3228&27 & 119 &144 & EK \\
\hspace{2mm}(HIDES; 68\,500; 3991--4815)  &  & & & & &  &  \\
 Th\"uringer Landessternwarte Tautenburg, Germany  &
2-m & 3142& 3219& 77& 215 & 116& HL\\
\hspace{2mm}(67\,000; 3700--5416) &  &  &  & & &  &  \\
La Silla Observatory, Chile &
1.2-m Euler & 3072& 3282& 211& 60 & 104& KU, KL, JV \\
\hspace{2mm}(CORALIE; 50\,000; 3876--6820) &   & & & & & &  \\
Mount John University Observatory, New Zealand  &
1.0-m &3154 &3193 & 40& 88 & 152& DW\\
\hspace{2mm}(HERCULES; 70\,000; 4456--7150)   &  & & & & &  &  \\
Observatorio Astron\'omico Nacional &  & & & & &  & \\
at San Pedro M\'artir, M\'exico & 2.1-m & 3205& 3211 &7 &637 & 282& RC, JE\\
\hspace{2mm}(Echelle Spectrograph; 20\,000; 3781--6893) &  &  & & & &  &  \\
\hline
  2007-2010   &&\multicolumn{2}{c}{$-$2450000} &&&&\\
\hline
Pic du Midi, France & 2.0-m TBL & 4286&5403 &1118  & 44 & 389& YF, CM, OT\\
\hspace{2mm}(NARVAL; 65\,000; 3694--10484)  &  & & & & &  &  \\
\hline
\hline
 Total&  &   & & 3674 & 1559 &  \\
\hline
\hline
\end{tabular}
\end{minipage}
\end{table*}


A total of 1354 high-resolution spectroscopic exposures was assembled using 10 different telescopes spread over both hemispheres. In addition, we added the 205 spectra of Neiner et al.\,(2003, 2012), which allowed us to increase the time span of the data set and, thus, to achieve a better frequency accuracy. Table\,\ref{TAB:observatories} summarizes the characteristics of the spectroscopic data. The resolving power $\lambda/\Delta\lambda$ of the spectrographs ranged from 20\,000 to 70\,000. When the instrument was not an \'echelle spectrograph, the spectral domain was chosen in order to cover the \SiIII triplet around 4567 \AA\ because these lines are strong without being much affected by blending. Moreover, they are dominated by temperature broadening so that the intrinsic profile can be modelled with a Gaussian, which simplifies the modelling of the line-profile variations for mode identification purposes.

All exposures were subjected to the usual reduction process, i.e., we applied debiasing, background subtraction, flat-fielding and wavelength calibration. Subsequently, the barycentric corrections were determined and the spectra were normalized to the continuum by fitting a cubic spline function. Special care was taken with regard to the normalisation as it is not only important for an abundance analysis but also for a line-profile variation study. Indeed, as pointed out in Zima\citet{zima06}, a mode identification with the FPF method (see Sect.\,\ref{SEC:MODEID}) is very sensitive to it.  

The average radial velocity (RV) per site is very similar for all spectra, so that there is no evidence that V2052~Oph is a spectroscopic binary. To correct for the slightly different zero points of the different telescopes, we proceeded in two different ways. In a first step, we shifted the spectra in such a way that the RV constant of a least-squares sine fit using the two dominant frequencies is put to the same value for each observatory. In another step, we shifted the spectra so that the minimum of the amplitude across the line profile for the radial mode (see Fig.\,\ref{radial}) is put at the same value for each dataset. The two approaches gave the same results in our further analysis.


\section{Atmospheric parameters and chemical composition}\label{sect_parameters}
Wolff \& Heasley\,\citet{wolff} estimated $T_{\rm eff}$=23\,000~K and $\log g$=4.2 based on Str\"omgren photometry and fitting of the H$\gamma$ line using Kurucz\,\citet{kurucz} atmospheric models, respectively. As shown by Nieva \& Przybilla\,\citet{nieva}, fitting the wings of the Balmer lines using LTE models systematically leads to an overestimation of the surface gravity. On the other hand, Niemczura \& Daszy\'nska-Daszkiewicz\,\citet{niemczura} obtained $T_{\rm eff}$=23\,350$\pm$650~K, $\log g$=3.89 and a metallicity of --0.25$\pm$0.16 dex with respect to solar based on LTE fitting of 
 {\it IUE} spectra. Using non-LTE TLUSTY models (Hubeny \& Lanz\,\citealp{hubeny}) with a solar helium abundance, Neiner et al.\,\citet{neiner03} found that $T_{\rm eff}$=21\,990~K and $\log g$=3.98 provided the best match to three \HeI lines, but that the Si lines were poorly fit in that case. Relaxing the constraint on the He content, they inferred $T_{\rm eff}$=25\,200$\pm$1100~K, $\log g$=4.20$\pm$0.11 and about a factor two helium overabundance with respect to solar (helium abundance by number, $y$$\sim$0.16). A satisfactory fit to the He and Si line profiles, as well as the continuum UV flux, was achieved in that case. 

Morel et al.\,\citet{morel06} used the non-LTE line-formation codes DETAIL/SURFACE (Butler \& Giddings\,\citealp{butler_giddings}; Giddings\,\citealp{giddings}) to carry out an EW-based analysis of the average of our 105 exposures from McDonald Observatory. Using a series of time-resolved spectra minimises the fact that the temperature varies significantly along the pulsation cycle ($\Delta T_{\rm eff}$$\sim$900~K; Morton \& Hansen\,\citealp{morton_hansen}; Kubiak \& Seggewiss\,\citealp{kubiak}). Fulfilling ionisation balance of \SiII/\SiIII and fitting the wings of four Balmer lines led to $T_{\rm eff}$=23\,000$\pm$1000~K and $\log g$=4.0$\pm$0.2. Guided by the results of Neiner et al.\,\citet{neiner03}, Kurucz models with He/H=0.178 were used. However, this choice has a negligible impact on the atmospheric parameters and abundances ($\Delta \log \epsilon$ $\la$ 0.05 dex). It is not possible to further constrain $T_{\rm eff}$ from ionisation balance of neon owing to the lack of measurable \NeII lines (Morel \& Butler\,\citealp{morel08}).

Sim\'on-D\'{\i}az\,\citet{simon_diaz} recently reported problems with the modelling of some Si lines using the non-LTE code FASTWIND (Puls et al.\,\citealp{puls}). 
The model atom used by Morel et al.\,\citet{morel06} is similar in many respects and one may expect similar problems to be encountered in our case for some of these lines which have been used to constrain $T_{\rm eff}$ (namely \SiII $\lambda$4128, which was the only \SiII line used, and \SiIII $\lambda$4813, 4829). We have therefore decided to redetermine $T_{\rm eff}$ using lines that are thought to be properly modelled only (i.e., \SiII $\lambda$6371 and \SiIII $\lambda$4568, 4575, 5740). As can be seen in Table\,\ref{tab_Si_abundances}, this leads to a much reduced abundance scatter and a mean value revised upwards to $\log \epsilon$(Si)=7.47$\pm$0.31 dex. More importantly, the Si ionisation balance is also satisfied in that case, thus supporting the previous estimate ($T_{\rm eff}$$\sim$23\,000~K).

\begin{table}
\centering
\caption{Equivalent widths (EWs) rounded off to the nearest m\AA \ and non-LTE abundances (on the scale in which $\log \epsilon$[H]=12) for the silicon lines properly modelled according to Sim\'on-D\'{\i}az\,\citet{simon_diaz} assuming $T_{\rm eff}$=23\,000~K, $\log g$=4.0 and a microturbulent velocity of 1 \kms.}
\label{tab_Si_abundances}
\begin{tabular}{lcc} \hline\hline
Transition                       & EW [m\AA] & $\log \epsilon$(Si) [dex]\\\hline
\SiII $\lambda$6371.4$^a$ & 43      &  7.47\\
\SiIII $\lambda$4567.8    & 100     &  7.38\\
\SiIII $\lambda$4574.8    & 66      &  7.39\\
\SiIII $\lambda$5739.7    & 85      &  7.65\\
\hline
Mean \SiII lines          &         &  7.47\\
Mean \SiIII lines         &         &  7.47\\
\hline
\end{tabular}
\begin{flushleft}
$^a$ Measured on the mean BOES spectrum, as this line falls at the extreme edge of an \'echelle order in the McDonald data and is hence unmeasurable. \\
\end{flushleft}
\end{table}

Neiner et al.\,\citet{neiner03} claimed V2052~Oph to be oxygen weak and helium strong based on spectral fitting using TLUSTY models. However, the O abundance they find ($\log \epsilon$=8.52$\pm$0.11 dex) may be regarded as typical of the values found for other B stars (see Morel\,\citealp{morel09} for a review). There is instead evidence for a nitrogen excess. The logarithmic CNO abundances ratios are [N/C]=--0.37$\pm$0.08 and [N/O]=--0.50$\pm$0.13 dex, and appear significantly higher than the solar values ([N/C]=--0.60$\pm$0.08 and [N/O]=--0.86$\pm$0.08 dex; Asplund et al.\,\citealp{asplund}). This is supported by the results of Morel et al.\,\citet{morel06} who found [N/C]=--0.22$\pm$0.19 and [N/O]=--0.40$\pm$0.35 dex.\footnote{These errors are the quadratic sum of the uncertainties in the CNO abundances. However, because the abundances have the same qualitative dependence against changes in the atmospheric parameters, computing the errors on the ratios themselves narrows down the errors and strengthens the case for an N excess: [N/C]=--0.22$\pm$0.18 and [N/O]=--0.40$\pm$0.22 dex.} A nitrogen overabundance is observed in other magnetic B stars on the main sequence. We refer to Morel\,\citet{morel10} and Martins et al.\,\citet{martins12} for a discussion on this matter.

Similarly to Neiner et al.\,\citet{neiner03}, a helium overabundance was reported by Morel et al.\,\citet{morel06}. However, the uncertainties were large and the evidence much weaker in that case ($y$=0.118$\pm$0.032). Although the helium-rich status of V2052~Oph has yet to be unambiguously established, this star stands out by the difficulty of fitting the lines using a single He abundance, a fact which explains the large error bars (Morel et al.\,\citealp{morel06}). This phenomenon might be related to surface inhomogeneities/vertical stratification of helium in this magnetic star. The recent spectropolarimetric measurements of Neiner et al.\,\cite{neiner12} showed that He patches are situated close to the magnetic poles.

\section{Line-profile study}\label{sect_lpv}
\subsection{Frequency analysis}\label{sect_fa}
For our line-profile study, we considered sufficiently strong and unblended lines of different chemical elements (He, C, N, O, Mg, Al, Si, S, Fe). These spectral lines were chosen among the list of lines used for the abundance analysis study by Morel et al.\,(2006; see their Table\,A.1). In particular, we analysed the \SiIII triplet around 4567 \AA\ and the \HeI line at 4713 \AA. The same periodicities were found whatever the considered lines. In what follows we only describe our frequency analysis on the \SiIII 4552 \AA\ line.
\subsubsection{Equivalent width and radial velocity data}

To perform our frequency analysis, we used the software package FAMIAS\footnote{FAMIAS has been developed in the framework of the FP6 European Coordination Action HELAS - \url{http://www.helas-eu.org/}} (Zima\,\citealp{zima08}). We started by examining the radial velocity (RV) and equivalent width (EW) measurements. In the RV data, we first recovered the known pulsation frequency $f_1$=$7.14846$~d$^{-1}$ and its harmonic $2 f_1$ with an amplitude of 6.7 and 1.5 km~s$^{-1}$, respectively. The error estimate of our determined frequency is between 0.0000002~d$^{-1}$ according to Montgomery \& O'Donoghue\,\cite{montgomery_donoghue} (calculated to be $\sigma_{f} = \sqrt(6) \sigma_{{\rm std}} / \pi \sqrt(N) A_{f} \Delta T$, where $\sigma_{{\rm std}}$ is the standard deviation of the final residuals, $A_{f}$ the amplitude of the frequency $f$, and $\Delta T$ the total timespan of the observations) and 0.0003~d$^{-1}$ (calculated to be 1/$\Delta T$). In order to be conservative, we adopt five significant digits, also for the other frequencies detected in the dataset. The EW data also vary with $f_1$. It is due to temperature variations caused by the oscillatory displacement of the stellar surface (De Ridder et al.\,\citealp{deridder02}). Such a behaviour is observed in several $\beta$~Cep stars with a high-amplitude mode, such as V1449~Aql (Briquet et al.\,\citealp{briquet09} and references therein). Besides $f_1$, we only found a second independent frequency in the RV and EW datasets: the frequency 0.27481~d$^{-1}$ and its harmonics. This variability was first discovered from the periodic variation of UV resonance lines observed with {\it IUE} (Neiner et al.\,\citealp{neiner03}, $P_{\rm rot}$=$3.638833$~d) and interpreted as due to the stellar rotation. This interpretation is definitely confirmed by recent magnetic field data that vary with the same period (Neiner et al.\,\citealp{neiner12}). Moreover, as we describe below, the observed behaviour in the spectral lines is typical of the presence of surface chemical inhomogeneities, which leads to line-profile variability as the star rotates. 

\begin{figure}
\vspace{-2.2cm}
\hspace{0.5cm}
\rotatebox{0}{\resizebox{0.95\columnwidth}{!}{\includegraphics{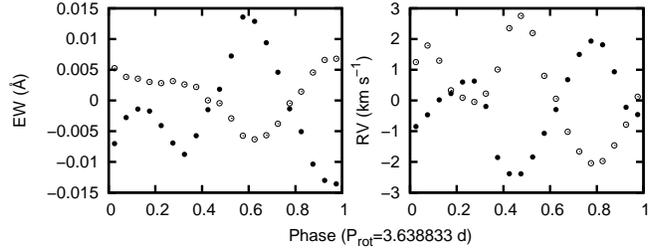}}}
\vspace{-6.1cm}
\caption{Phase diagrams of the phase-binned average EW (left) and RV (right) measurements computed from the \HeI\ 4713\ \AA\ line (full circles) and the \SiII\ 4552 \AA\ line (open circles), after prewhitening the pulsation frequency $f_1$, for $P_{\rm rot}$.} 
\label{ew_rv}
\end{figure}

In Fig.\,\ref{ew_rv}, we show phase diagrams of the EW and RV measurements after prewhitening the pulsation frequency $f_1$, for $P_{\rm rot}$. These plots resemble those of the B6 star HD\,105382 and of the B7 star HD\,131120 (see Briquet et al.\,\citealp{briquet01}, \citealp{briquet04}). These two stars are He-weak stars whose monoperiodic light and line-profile variability is interpreted as due to surface chemical spots, which are the result of complex interactions between radiatively driven diffusion processes and a stellar magnetic field (Alecian et al.\,\citealp{alecian11}). For V2052~Oph, Neiner et al.\,\citet{neiner03} concluded that the observed rotational modulation is also mainly due to chemical variations, as the weak magnetic field of the star is unlikely to produce strong temperature effects (see discussion in their Sect.\,7.1). As shown in Briquet et al.\,\citet{briquet01}, a non-sinusoidal RV variation with $f_{\rm rot}$ and $2f_{\rm rot}$ can be reproduced by two spots of the same element having a longitude difference of $\sim$180$^\circ$ to account for the two maxima (minima) separated by a phase difference of $\sim$0.5. Moreover, the out of phase variation of the EW and RV data can be explained if ionized helium is enhanced in regions of the stellar surface where \SiII is depleted and vice versa. This trend was found for both HD\,105382 and HD\,131120 by deriving detailed abundance maps for Si and He (Briquet et al.\,\citealp{briquet04}). By analogy, the same conclusion may be obtained for V2052~Oph. An inversion of the observed rotationally modulated line-profile variations into a two-dimensional abundance distribution (so called Doppler mapping; e.g. Piskunov \& Rice\,\citealp{piskunov93}) could confirm it but is beyond the scope of this paper. In the case of V2052~Oph, the difficulty is to disentangle the variability due to pulsation and the one due to rotational modulation that is also found in spectral lines of other chemical elements such as C, N, O and Fe. Moreover, while diffusion models in magnetic Ap stars are available for a comparison with observed abundance maps (Michaud et al.\,\citealp{michaud81}; Vauclair et al.\,\citealp{vauclair91}), such theoretical models need to be developed for stars as hot as V2052~Oph.


\begin{figure}
\vspace{5.4cm}
\rotatebox{-90}{\resizebox{0.72\columnwidth}{!}{\includegraphics{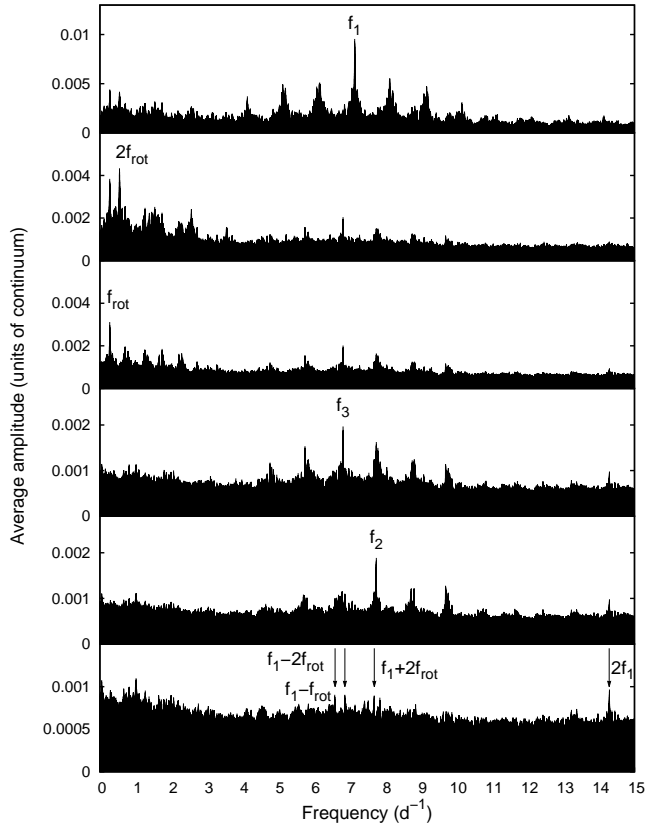}}}
\caption{Amplitude periodogram, which is an average of all Fourier spectra across the \SiIII 4552 \AA\ profile. The different panels show the periodograms in successive stages of prewhitening (top to bottom). We follow the notation of Handler et al.\,\citet{handler12} for the frequencies. Note the different scale of the Y-axis in the different panels.
}

\label{periodo}
\end{figure}

\subsubsection{2D analysis}
After a one-dimensional frequency search on the integrated quantities, we performed a pixel-by-pixel Fourier analysis across the line profile (2D analysis) by means of the tools available in FAMIAS. In Fig.\,\ref{periodo}, the mean of all Fourier spectra across the \SiIII 4552 \AA\ line is shown for subsequent steps of prewhitening. The significance of a frequency cannot be derived from the average Fourier spectrum but one has to proceed as follows. The wavelength at which the given frequency has the highest amplitude is determined. Then, the Fourier spectrum at this pixel is computed and the frequency is retained if its amplitude exceeds 4 times the signal-to-noise (Breger et al.\,\citealp{breger93}). The noise level is calculated in a 5\,d$^{-1}$ interval centered on the frequency of interest. As for some other $\beta$~Cep stars (Telting et al.\,\citealp{telting97}; Schrijvers, Telting \& Aerts\,\citealp{schrijvers04}; Briquet et al.\,\citealp{briquet05}), the pixel-by-pixel analysis led to additional independent frequencies compared to the 1D frequency search: $f_2$=$7.75603\ \rm{d}^{-1}$ and $f_3$=$6.82308\ \rm{d}^{-1}$. In the residual periodogram, the frequency $2f_1$ and combinations of the dominant frequencies $f_1$, $f_{\rm rot}$ and $2f_{\rm rot}$ are also seen, as expected in a 2D search (Zima\,\citealp{zima08}). Because the 2D analysis is more sensitive to the detection of high-degree modes than a 1D analysis, it indicates that $f_2$ and $f_3$ are probably not low-degree modes. The four frequencies found in our spectra were also discovered from the multisite photometry by Handler et al.\,\citet{handler12} and are thus certain.

\subsection{Mode identification}\label{SEC:MODEID}

\subsubsection{Previous results}\label{sec:previous}

The dominant mode with frequency $f_1$=$7.14846$ d$^{-1}$ was first identified as radial by Heynderickx, Waelkens \& Smeyers\,\citet{heynderickx04} and Cugier, Dziembowski \& Pamyatnykh\,\citet{cugier94}. This is unambiguously confirmed by the multisite photometric campaign presented in Handler et al.\,\citet{handler12}. As for the frequencies $f_2$=$7.75603\ \rm{d}^{-1}$ and $f_3$=$6.82308\ \rm{d}^{-1}$, Handler et al.\,\citet{handler12} deduced that both modes correspond to $\ell$=4 or $\ell$=6 by comparing the $uvy$ passband amplitudes with the theoretically predicted ones. The mode with frequency $f_3$ was first identified by Neiner et al.\,\citet{neiner03} as an $\ell_3$=3 or 4 mode by means of the MUSICOS spectra. A study by Pollard et al.\,\citet{pollard08} of the HERCULES spectra also revealed this mode. Their mode identification showed that it corresponds to an $\ell_3$=4$\pm$1 and $m_3$=2$\pm$2 with a best fitting obtained for $(\ell_3,m_3)$=$(4,2)$. The combination of these photometric and spectroscopic results allows us to safely conclude that $\ell_3$=$4$ for $f_3$. Because of geometrical cancellation, Handler et al.\,\citet{handler12} concluded that the most likely identification for $f_2$ is also $\ell_2$=$4$. We check these conclusions through our mode identification aimed at giving constraints on $(\ell,m)$ for $f_2$ and $f_3$ (see Sect.\,\ref{sec:mi}).

\begin{figure}
\centering
\vspace{-0.7cm}
\begin{center}
 \begin{tabular}{cc}
\resizebox{0.4\textwidth}{!}{\includegraphics{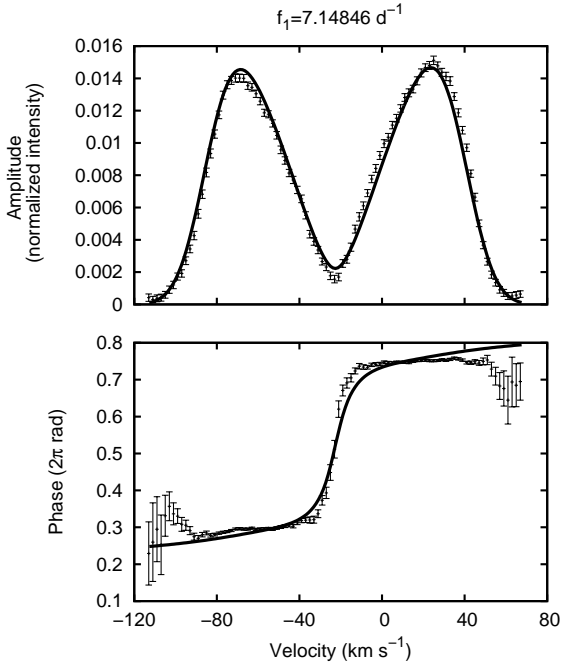}}
\end{tabular}
\end{center}
\vspace{-1cm}
\caption{
Observed amplitude and phase distributions of the $f_1$ frequency across the \SiIII 4552 \AA\ line profile (points with error bars) compared to the theoretical ones (full line) for $(\ell_1,m_1)$=$(0,0)$. The amplitude is expressed in units of continuum and the phase in $2\pi$ radians.}
\label{radial}
\end{figure}

\subsubsection{The method used}
Two independent methods based on spectroscopy are usually used to identify the modes of main-sequence pulsators hotter than the Sun, namely the moment method (Briquet \& Aerts\,\citealp{briquet03}) and the Fourier parameter fit (FPF) method (Zima\,\citealp{zima06}). Both methods are implemented in the software package FAMIAS and a typical application of them on a $\beta$~Cep star can be found in Desmet et al.\,\citet{desmet09} for 12~Lac. For V2052~Oph, $f_2$ and $f_3$ are detected in the second moment (line width) but not in the first moment (radial velocity). Therefore, we only made use of the FPF method, which is appropriate since the projected rotation velocity of our target is sufficiently large ($ v\sin i > 20\ {\rm km\ s} ^{-1}$). 

In the FPF technique, a multi-periodic non-linear least-squares fit of sinusoids is computed using all detected frequencies for every wavelength bin across the profile according to the formula $Z+\sum_i A_i \sin\bigl[2\pi (f_i t+\phi_i)\bigr],$ where $ Z$ is the zero-point, and $ A_i$, $ f_i$, and $ \phi_i$ are the amplitude, frequency, and phase of the $i$-th frequency, respectively. This gives, for each pulsation frequency, the observational values of the zero-point, amplitude and phase as a function of the position in the line profile. The wavenumbers $(\ell,m)$ and other parameters, such as the stellar inclination angle $i$, are then determined in such a way that the theoretically computed zero-point, amplitude and phase values derived from synthetic line profiles best fit the observed values. The goodness of the fit is expressed as a reduced $ \chi^2$-value (see Zima\,\citealp{zima06}). The approximations made in FAMIAS to compute the synthetic profiles (see Zima\,\citealp{zima06} for details) remain valid for our study case. As for stellar rotation, the first order effects of the Coriolis force are taken into account in the displacement field.

\subsubsection{Additional constraints}\label{sec:mi}


\begin{table}
\centering
  \caption{The best-fit solutions of our mode identification with the FPF method for $f_2$=$7.75603$ d$^{-1}$ and $f_3$=$6.82308$ d$^{-1}$. $f_1$=$7.14846\ \rm{d}^{-1}$ is adopted as the radial mode. $i$ and $v_{\rm eq}$ are the stellar inclination angle and the equatorial rotational velocity, respectively.
}
\label{table_mi}
  \begin{tabular}{@{}lcccccc}
\hline 
ID & $\chi^2$ & ($\ell_2,m_2$) & ($\ell_3,m_3$) & $i$ & $v_{\rm eq}$ \\
& &  &  & ($^{\circ}$) & (km~s$^{-1}$)  \\
\hline
1 & 1.16 &  $(4,3)$ & $(4,2)$ & 59.7$\pm$2.1 &   75.0$\pm$1.6 \\
2 & 1.23 &  $(4,3)$ & $(4,4)$ & 33.6$\pm$1.9  &  118.2$\pm$5.1 \\
3 & 1.31 &  $(4,3)$ & $(4,3)$ & 81.6$\pm$0.7  &  65.3$\pm$0.5  \\ 
4 & 1.42 &  $(4,2)$ & $(4,2)$ & 53.9$\pm$3.1  &  80.2$\pm$3.5  \\
5 & 1.57 &  $(4,2)$ & $(4,4)$ & 46.9$\pm$2.4  &  89.1$\pm$3.4   \\
6 & 1.63 &  $(4,2)$ & $(4,3)$ & 55.3$\pm$7.9  &  78.0$\pm$7.7  \\
7 & 1.81 &  $(4,4)$ & $(4,2)$ & 45.8$\pm$2.6  &  90.5$\pm$3.5   \\
8 & 1.86 &  $(4,4)$ & $(4,3)$ & 28.5$\pm$3.3  &  136.1$\pm$14.9 \\
9 & 2.01 &  $(4,4)$ & $(4,4)$ & 27.5$\pm$1.1 &   141.2$\pm$5.5   \\
\hline
\end{tabular}
\end{table}

\begin{figure}
\vspace{-2.4cm}
\rotatebox{0}{\resizebox{\columnwidth}{!}{\includegraphics{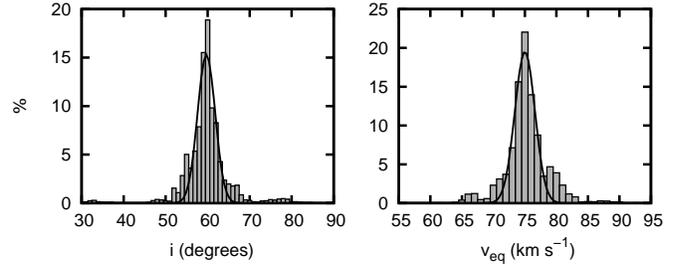}}}
\vspace{-6.5cm}
\caption{Histograms for the inclination and equatorial rotational velocity of the star derived from the FPF method, for the combination $(\ell_2,m_2)$=$(4,3)$ and $(\ell_3,m_3)$=$(4,2)$.} 
\label{histo}
\end{figure}

\begin{figure*}
\centering
\begin{center}
 \begin{tabular}{cc}
\vspace*{-0.5cm}
\resizebox{0.45\textwidth}{!}{\includegraphics{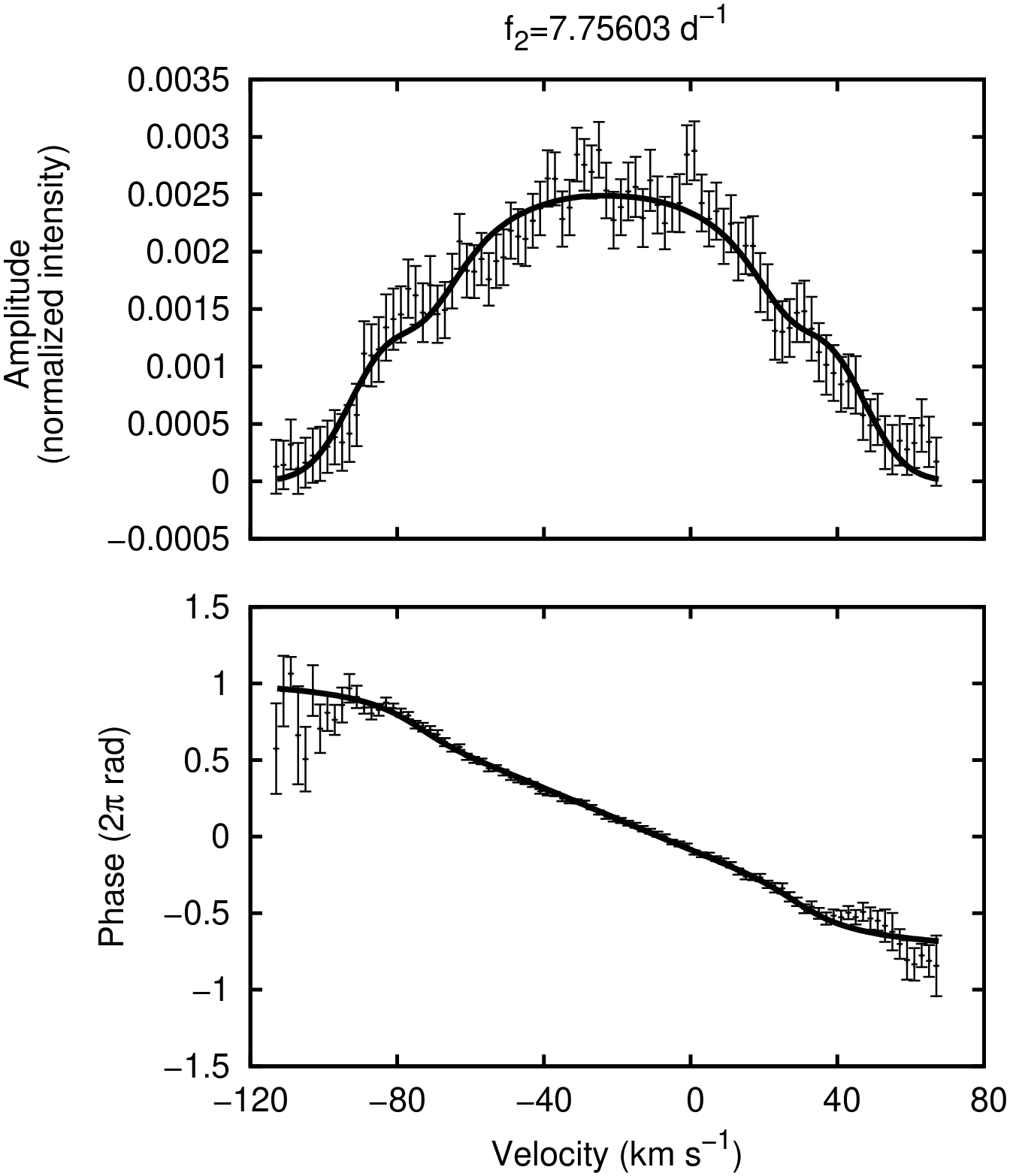}}&
\resizebox{0.45\textwidth}{!}{\includegraphics{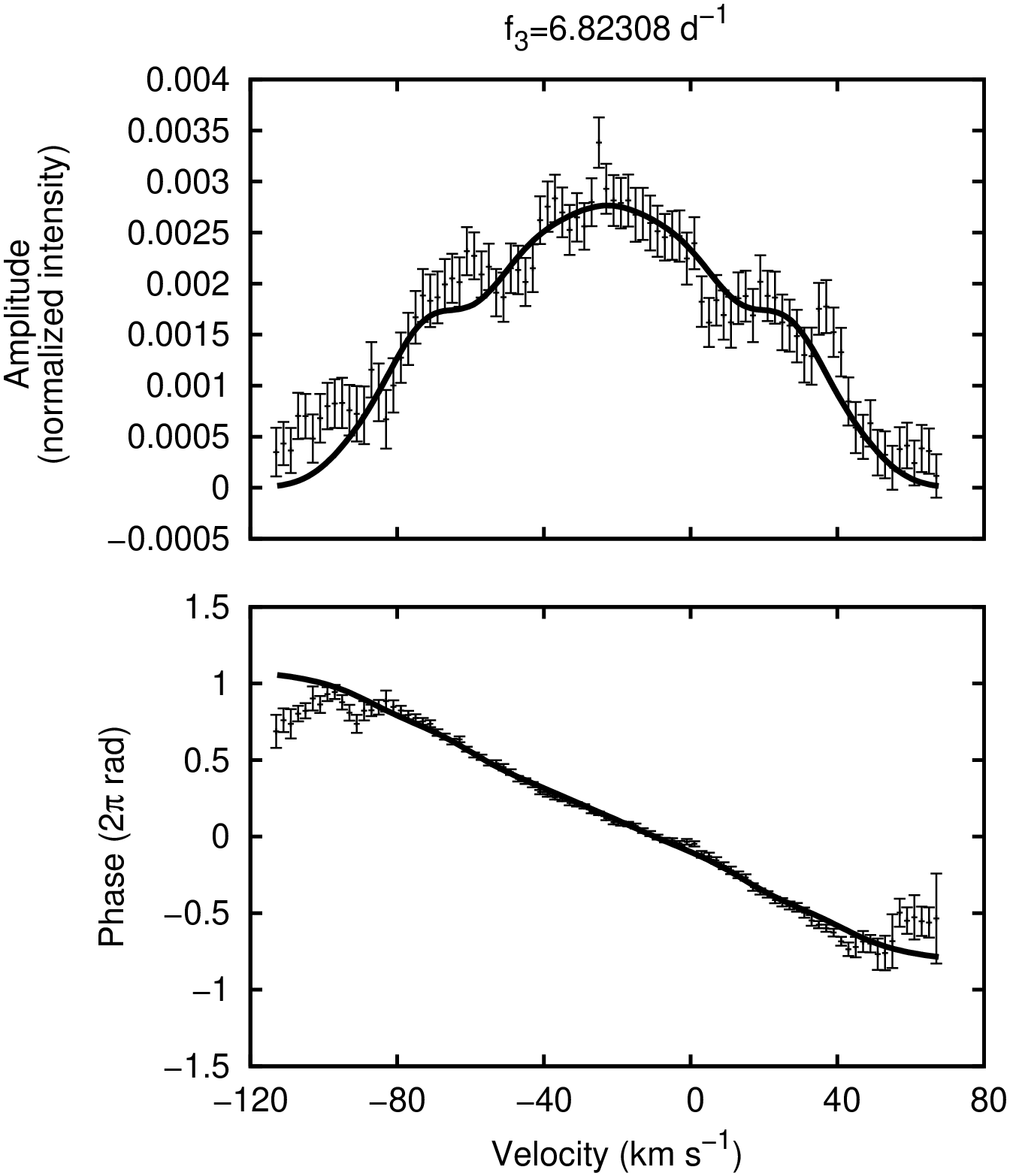}}\\
\end{tabular}
\end{center}
\vspace*{-5mm}
\caption{Amplitude and phase distributions for $f_2$ and $f_3$ for the \SiIII 4552 \AA\ line (points with error bars) and comparison with the theoretical ones (full lines) for the solution with $(\ell_2,m_2)$=$(4,3)$ and $(\ell_3,m_3)$=$(4,2)$. The amplitudes are expressed in units of continuum and the phases in $2\pi$ radians. }
\label{amp_phase}
\end{figure*}

We applied the FPF method using the FAMIAS software to identify the modes detected in our spectroscopic data. In our procedure, we always adopted $(\ell_1,m_1)$=$0$ for the dominant mode as deduced from photometry and confirmed by our spectroscopy. Indeed, the amplitude and phase behaviour across the line profile for this mode is very typical for a radial pulsation mode (see Fig.\,\ref{radial}).

We adopted the values $M=9\, M_\odot$ and $R=5\,R_\odot$ for the mass and radius, respectively (Handler et al.\,\citealp{handler12}). The mode identification results are robust when using different values for these parameters, within the errors. Before starting the mode identification, the parameter space was restricted by estimating values for the projected rotational velocity $v \sin i$, the width of the intrinsic gaussian profile $\sigma$, and the equivalent width $EW$ from a least-squares fit of a rotationally broadened synthetic profile to the zero-point profile. A genetic optimization with the following free parameters was afterwards adopted: $\ell \in [0,8]$ with a step of 1, $m \in [-\ell,\ell]$ with a step of 1, the surface velocity amplitude $a \in [1,50]\,\kms$ with a step of 1\,\kms, the stellar inclination angle $i \in [1,90]\,\deg$ with a step of $1\,\deg$, $v \sin i \in [50,70]\,\kms$ with a step of 0.1\,\kms and $\sigma \in [5,15]\,\kms$ with a step of 0.1\,\kms. The fitting between the observed and theoretical zero-point, amplitude and phase was carried out by applying genetic optimization routines. First, we used mono-mode fits in order to speed up the computations. Once constraints on the parameters are obtained, we performed a multi-mode fit.

The results clearly exclude $\ell$ below 3 and above 5 for both low-amplitude modes. Combining this outcome with the photometric one ($\ell$=4 or 6; Handler et al.\,\citealp{handler12}), we find that $\ell$=$4$ for $f_2$ and $f_3$. In order to fit the phase across the profile, we also found that $m > 0$ (where a positive $m$-value denotes a prograde mode) for both modes. Moreover, a visual inspection of the solutions revealed that we could also exclude the ($\ell,m$)-combinations involving $m$=$1$. The matching between the observed and synthetic amplitudes across the profile was not satisfactory in these cases. 

In Table\,\ref{table_mi}, we list the remaining nine acceptable ($\ell,m$)-combinations and give estimates of the stellar inclination angle $i$ and equatorial rotational velocity $v_{\rm eq}$ obtained through a simultaneous fit of the modes. Histograms, constructed using the $\chi^2$-value as a weight, for the solution with the lowest $\chi^2$ (see Table\,\ref{table_mi}) are displayed in Fig.\,\ref{histo}. Fig.\,\ref{amp_phase} depicts the matching between the observed and theoretical amplitude and phase across the line, for the combination $(\ell_2,m_2)$=$(4,3)$ and $(\ell_3,m_3)$=$(4,2)$. 

\section{Comparison with stellar models}\label{sect_modelling}

We computed stellar models accounting for the derived observational properties of V2052~Oph. To this end, we used the evolutionary code CL\'ES (Code Li\'egeois d'\'Evolution Stellaire; Scuflaire et al.\,\citealp{scuflaire08a}) with the input physics described in Briquet et al.\,\citet{briquet11}. For each stellar model considered, the theoretical pulsation frequency spectrum was calculated in the adiabatic approximation with the code LOSC (Scuflaire et al.\,\citealp{scuflaire08b}). Once models that fit the observed modes were selected, we checked the excitation of the pulsation modes with the linear non-adiabatic code {\sc MAD} (Dupret\,\citealp{dupret01}; Dupret et al.\,\citealp{dupret02}). In our computations, the effects of rotation on the pulsation frequencies were taken into account to first order, which is sufficient for our approach as justified {\it a posteriori} at the end of this section.

By varying the mass ($M$ $\in$ [7.6, 20.0] M$_\odot$ with a step of 0.1 M$_\odot$), the hydrogen mass fraction ($X$ $\in$ [0.68, 0.74] with a step of 0.02), the metallicity ($Z$ $\in$ [0.010, 0.018] with a step of 0.002) and the core overshooting parameter ($\alpha_{\rm ov}$ $\in$ [0, 0.5] with a step of 0.05 -- expressed in local pressure scale heights), we computed models from the zero-age main-sequence (ZAMS) to the terminal-age main-sequence (TAMS).

Given the rotation period of the star and the model radius when $f_1$ is identified as the fundamental, we get an equatorial rotational velocity between 71 and 75 km~s$^{-1}$ for V2052~Oph. With $v \sin i \in [61,65]$ km~s$^{-1}$, the star has an inclination $i \in [54,66]$$^{\circ}$. This range of inclination angles is in full agreement with the one determined by Neiner et al.\,\citet{neiner12} who obtained an inclination between 53$^{\circ}$ and 77$^{\circ}$ by synthesizing the Stokes V profiles of the star with a centered or offcentered dipole model. Stellar models with the radial mode as the first overtone (or higher overtones) have larger radii, implying an inclination lower than 50$^{\circ}$ that is not compatible with the constraints deduced from the magnetic field data. We refer to Hander et al.\,\cite{handler12} for a deeper discussion.

\begin{figure}
\centering
\includegraphics[angle=270,totalheight=0.33\textwidth]{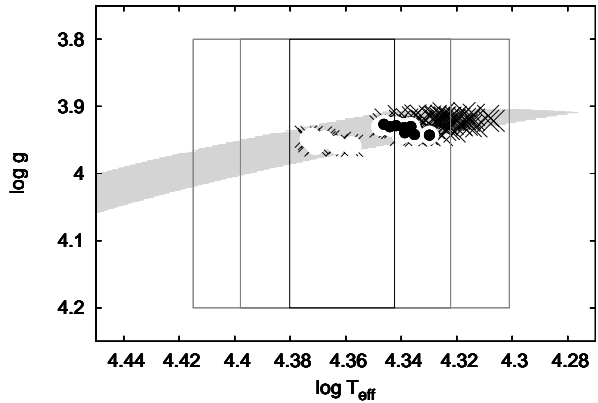}
\caption{The error box represents the position of V2052~Oph, deduced from spectroscopy, in the $\log T_\mathrm{eff}$--$\log g$ diagram. The 1, 2 and 3-$\sigma$ regions in $T_\mathrm{eff}$ are shown. The light grey band corresponds to models that fit the radial fundamental mode. Our best models for the star are displayed as crosses (considering all nine ($\ell,m$)-combinations of Table\,\ref{table_model}), white circles (with all solutions but ID 7, 8 and 9), black circles (for the solution with ID 1). We refer to the text for more explanations.}
\label{fig_model}
\end{figure}

From the models that we computed, we thus selected the ones along the main-sequence evolutionary track that fit the radial fundamental mode to within 0.05 d$^{-1}$. The position of these models in the $\log T_\mathrm{eff}$--$\log g$ diagram is compared to the spectroscopic error box of the star in Fig.\,\ref{fig_model}. For all models matching $f_1$, we computed the theoretical frequency spectrum of low-order $p$- and $g$-modes with a degree of $\ell$=$4$. The rotationally split frequencies were derived using the Ledoux constant as $f_{n \ell m}=f_{n \ell 0}+m \beta_{n \ell} f_{\rm rot}$, where $\beta_{n \ell}$ is a structure constant depending on the stellar model. Afterwards, to each model fitting the radial fundamental mode, we assigned a $\chi^2$-value, which compares the observed frequencies $f_i^{\rm obs}$ with the theoretically computed ones $f_i^{\rm th}$, as follows:
\begin{equation}\label{eq}
\centering
\chi_f^2=\sum_{i=2}^{3}\frac{(f_i^{\rm obs}-f_i^{\rm th})^2}{\sigma^2}
\end{equation}
where $\sigma$ is taken to be 0.05 d$^{-1}$ to account for uncertainties in our modelling prescriptions (see last paragraph of this section). The models corresponding to one of the nine ($\ell,m$)-combinations of Table\,\ref{table_mi} and having a $\chi_f^2 < 1$ were retained, also requiring the excitation of the radial mode. These models are displayed as crosses in Fig.\,\ref{fig_model} and their physical parameters are listed in Table\,\ref{table_model}.

\begin{table*}
\centering
  \caption{Physical parameters of the models fitting the three pulsation modes detected in V2052~Oph according to their mode identifications listed in Table\,\ref{table_mi}, using the same IDs. The last line corresponds to the parameter range of our best models for the star, i.e., with a $\chi_f^2$ as defined in Eq.~(\ref{eq}) lower than 1 and excluding solutions with ID 7, 8 and 9.
}
\label{table_model}
  \begin{tabular}{@{}lcccccccccc}
\hline 
ID & mass & radius & Z & $\alpha_{ov}$& $T_{\rm eff}$ &$\log g$ & $\log (L/{\rm L}_\odot)$ & age & $X_c$ \\
 & (M$_\odot$) & (R$_\odot$) & & & (K) &  & & (Myr) & \\
\hline
1 &  [8.2, 9.0] & [5.16, 5.31] & [0.012, 0.016] & [0.00, 0.15] & [21370, 22190] & [3.93, 3.94] & [3.72, 3.76] & [19.9, 23.7]  & [0.28, 0.32]\\
2 & no model & & & & & & & & \\ 
3 &  [8.9, 9.6] & [5.26, 5.38] & [0.010, 0.012] & [0.00, 0.10]  & [22790, 23570] & [3.95, 3.96] & [3.85, 3.89] &  [17.3, 18.3] & [0.27, 0.31]\\
4 &  [8.3, 9.0] & [5.17, 5.30] & [0.012, 0.016] & [0.00, 0.10]  & [21600, 22250] & [3.93, 3.94] & [3.73, 3.77] &  [19.2, 21.9] & [0.27, 0.30]\\
5 & no model & & & & & & & & \\
6 & [9.0, 9.1] & [5.27, 5.29] &      0.010   &    0.05  &  [23320, 23620]     &   3.95    &  [3.89, 3.87] &     [16.9, 18.0]   &   [0.25, 0.27]\\
7 & [8.1, 9.0] & [5.15, 5.31] &  [0.016, 0.018] &  [0.00, 0.15] & [20550, 21720] & [3.92, 3.94] & [3.64, 3.74] & [18.7, 26.5] & [0.30, 0.35]\\
8 & [7.8, 9.6] & [5.10, 5.39] &  [0.010, 0.018] &  [0.00, 0.35] & [20260, 23680] & [3.91, 3.96] & [3.61, 3.90] & [15.6, 28.9] & [0.28, 0.37]\\
9 & [7.7, 8.2] & [5.10, 5.19] &  [0.014, 0.018] &  [0.25, 0.40] & [20350, 21620] & [3.91, 3.92] & [3.62, 3.71] & [24.5, 30.1] & [0.33, 0.38]\\
\hline
 & [8.2, 9.6] & [5.16, 5.38] &  [0.010, 0.016] &  [0.00, 0.15] & [21370, 23620] & [3.93, 3.96] & [3.72, 3.89] & [16.9, 23.7] & [0.25, 0.32]\\
\hline
\end{tabular}
\end{table*}

Although we permitted the two $\ell$=$4$ modes to have the same radial overtone, none of the models matching our observations are such that $f_2$ and $f_3$ belong to the same rotationally split multiplet. In all models appropriate for the star, the frequencies $f_2$ and $f_3$ have $n_2$=$-3$ ($g_3$-mode) and $n_3$=$-2$ ($g_2$-mode), respectively. Both frequencies are predicted to be excited by non-adiabatic computations in all cases. In fact, a scan of the modes with $\ell$=$4$ shows that only those with $n$=$-1$,\ $-2$, and $-3$ are theoretically excited. Following theory, modes with $n$=$-1,\ 1$ for $\ell$=$1$, $n$=$-1,\ -2$ for $\ell$=$2$, and $n$=$-1,\ -2,\ -3$ for $\ell$=$3$ are also excited in the same frequency range as the observed modes. Clearly, if they are present in V2052~Oph, their amplitudes are lower than the observed $\ell$=$4$ modes. It may be explained by the rotation rate of the star that is somewhat higher than in other well-studied $\beta$~Cep stars with dominant low-degree modes (see Hander et al.\,\citealp{handler12} for more comments).

In Table\,\ref{table_mi}, the inclination angles $i$ of solutions with ID 1, 4 and 6 are fully compatible with the stellar modelling. Solutions with ID 2 and 5 are rejected because no model with the corresponding ($\ell,m$)-combinations could be found (see Table\,\ref{table_model}). If one limits the inclination range around 60$^{\circ}$ during the mode identification as performed in Sect.\,\ref{SEC:MODEID}, solutions with ID 7, 8 and 9 have even higher $\chi^2$-values so that we can exclude them. However, solution with ID 3 is kept since the $\chi^2$-value is similar as before (1.33). The most probable identification for both $f_2$ and $f_3$ is thus $(\ell,m)$=$(4,2)$ or $(4,3)$.

The parameter ranges of our best models for V2052~Oph are listed in the last line of Table\,\ref{table_model}. They are obtained by only considering the solutions with ID 1, 3, 4 and 6 and are represented as white circles in Fig.\,\ref{fig_model}. In this figure, we also distinguish the models associated to ID 1 as black circles. An important result is that only stellar models with no or mild core overshooting can explain our observational constraints. This is further discussed in the following section. One also notices the excellent agreement between our best fitting model parameters and the $T_{\rm eff}$ and $\log g$-values derived from spectroscopy. Among $\beta$~Cep stars with an asteroseismic modelling, such a consistency was also obtained for 12~Lac (Desmet et al.\,\citealp{desmet09}) but it is not often the case (e.g. Briquet et al.\,\citealp{briquet07}; Briquet et al.\,\citealp{briquet11}; Aerts et al.\,\citealp{aerts11}). 

In order to assess the influence of rotation on the pulsation frequency values, we calculated, using the 2D method described in Reese, Ligni\`eres \& Rieutord\,\citet{reese06}, the fundamental radial mode and the $\ell$=$4$, $n$=$-2,\,-3$ multiplets of rotating polytropic models with a polytropic index of $3$ (which is typical of a radiative zone), and a mass and polar radius set to $9\,{\rm M}_{\odot}$ and $5\,{\rm R}_{\odot}$, respectively.  A comparison between a first order approximation of the frequencies and the 2D calculations yielded differences around $0.05\,\mathrm{d}^{-1}$ for the radial mode, and $0.02\,\mathrm{d}^{-1}$ or less for the $\ell$=$4$ modes, for an equatorial rotational velocity of $75\,\mathrm{km\,s}^{-1}$.  As explained in Ballot et al.\,\citet{ballot10}, $p$-modes are more affected by the centrifugal deformation, given that their energy is concentrated towards the outer layers of the star, thereby explaining the larger difference for the radial fundamental mode. 
Although these differences can be detected, given the length of our observational run, they remain below the percent level compared to the frequencies. Furthermore, they are much smaller than the distance between the modes $(n,\,\ell,\,m)=(-2,4,0)$ and $(-3,4,0)$, and the spacing between consecutive members of the $g$-mode multiplets when the frequencies are expressed in an inertial frame. The above differences explain why we considered an error bar of $0.05\,\mathrm{d}^{-1}$ on the pulsation frequencies for our comparisons between observed and model frequencies.

\section{Discussion}\label{sect_discussion}

V2052~Oph is very similar in terms of stellar parameters to the $\beta$ Cep star $\theta$~Oph, except that V2052~Oph has a higher surface rotation velocity (see Table\,\ref{compar}). $\theta$~Oph rotates at 6\% of its critical breakup velocity, while V2052~Oph rotates at 16\%. For such stars, we expect a larger core overshooting parameter for the faster rotator. Indeed, Browning et al.\,\citet{browning04} showed that when the rotation rate is increased (for given stratification and diffusivities) the penetration of flows coming from the convective core in the surrounding stably-stratified radiative envelope becomes more intense because of the stabilisation of the flow by the Coriolis acceleration. In addition, the rotational mixing associated to the possible differential rotation increases with rotation rate (Mathis \& Zahn\,\citealp{mathis04}; Maeder\,\citealp{maeder09}). 

The reverse result is obtained since the core overshooting parameter is larger for $\theta$~Oph than for V2052~Oph. Indeed, Briquet et al.\,\citet{briquet07} obtained $\alpha_{\rm ov}$$=$0.44$\pm$0.07 $H_p$ with $Z$=$0.009$ to 0.015 for $\theta$~Oph, using the same tools and input physics as here. Using a different stellar evolution modelling code and a 2D linear adiabatic pulsation code, Lovekin \& Goupil\,\citet{lovekin10} found $\alpha_{\rm ov}$=0.28$\pm$0.05 $H_p$ assuming $Z$=$0.02$. The results of both studies are compatible since increasing $Z$ decreases $\alpha_{\rm ov}$, as shown in Briquet et al.\,\citet{briquet07}. Such values were also found in the modelling by Daszy\'nska-Daszkiewicz \& Walczak\,\citet{daszy09}.

The inhibition of mixing as derived for V2052~Oph can be caused by processes that are able to damp differential rotation in the stellar radiation zone, i.e. either stochastically excited internal waves (e.g. Talon \& Charbonnel\,\citealp{talon05}; Pantillon, Talon \& Charbonnel\,\citealp{pantillon07}) or a magnetic field (e.g. Moss\,\citealp{moss92}). The relative size of the convective core of the two stars is similar: 12.5\% of the total radius for V2052~Oph and 13.5\% for $\theta$~Oph. Therefore, if internal waves transporting angular momentum are excited by the turbulent convection in the core, they are expected to have the same net effect in the two stars. V2052~Oph is known to be a magnetic star (Neiner et al.\,\citealp{neiner03}) while $\theta$~Oph is not (Hubrig et al.\,\citealp{hubrig06}). V2052~Oph hosts a fossil magnetic field with $B_{\rm pol}$ of the order of 400~G (Neiner et al.\,\citealp{neiner12}). Therefore, the presence of the magnetic field in V2052~Oph seems to be the most likely explanation for the low core overshooting parameter, which represents the non-standard mixing processes. We test below whether the observed field can be sufficient to inhibit mixing in V2052~Oph.

\begin{table}
\caption[]{Comparison of the parameters of V2052~Oph and $\theta$~Oph.}
\begin{center}
\begin{tabular}{lll}
\hline
\hline
Parameter 		& $\theta$~Oph		& V2052~Oph 		\\
\hline
$T_{\rm eff}$ (K)	        & 22260$\pm$280	        & 22500$\pm$1100	\\
$\log g$ (dex)		& 3.95$\pm$0.01  	& 3.95$\pm$0.02		\\
$M$ (M$_\odot$)		& 8.2$\pm$0.3		& 8.9$\pm$0.7		\\
$X_c$			& 0.38$\pm$0.02	        & 0.29$\pm$0.04 	\\
$Z$			& [0.009, 0.015]		& [0.010, 0.016]		\\
$\alpha_{\rm ov}$		& 0.44$\pm$0.07	        & 0.07$\pm$0.08 	\\
$v_{\rm eq}$ (km~s$^{-1}$)	& 29$\pm$7		& 73$\pm$2		\\
$\Omega / \Omega_{\rm crit}$ & 0.09$\pm$0.03 & 0.23$\pm$0.01			\\
\hline
\end{tabular}
\begin{flushleft}
The parameters for $\theta$~Oph are taken from Briquet et al.\,\citet{briquet07}.
\end{flushleft}
\end{center}
\label{compar}
\end{table}

We first used the criterion from Eq.~(3.6) in Zahn\,\citet{zahn11}, which is based on the equation of angular momentum transport taking into account the magnetic field derived by Mathis \& Zahn\,\citet{mathis05}:
\begin{equation}
B_{\rm crit} ^2 =4 \pi \rho \, {R^2 \Omega \over t_{\rm AM}}.
\end{equation} 
In this equation, we used the time already spent by V2052~Oph on the main sequence, its radius and mass, as well as the surface rotation velocity. We found that the mean critical field strength in the radiative zone for suppressing differential rotation in this star is $B_{\rm crit}$$\sim$70~G. According to the ratio (30) between internal and surface fields derived by Braithwaite (2008; see Fig.\,8 therein), this corresponds to a critical field strength at the surface of $B_{\rm crit,surf}$=$2$~G. Although the Zahn criterion is an approximation and we took global parameter values in our numerical application, it provides a good idea that a very weak fossil magnetic field is enough to inhibit possible initial differential rotation and thus rotational mixing in V2052~Oph or similar early B stars. Thus the surface magnetic field of 400~G present in V2052~Oph is strong enough and V2052~Oph is rotating uniformly considering the approximations of the Zahn criterion.

We then considered the criterion from Eq.~(22) defined by Spruit\,\citet{spruit99}, which provides the critical initial field strength above which the magnetic field remains non-axisymmetric and rotation becomes uniform:
\begin{equation}
B_1=r(4\pi\rho)^{1/2} \left({\eta\Omega^2q^2\over 3r^2\pi^2}\right)^{1/3}.
\end{equation} 
In this equation we consider the radiative zone. From our CL\'ES models we derived that this radiative zone encompasses 81\% of the stellar mass and its average density is 0.07 g~cm$^{-3}$. In addition, we assumed that the initial rotational velocity in the radiative zone is a few times the current surface rotation velocity. We used a magnetic diffusivity of $2 \times 10^6$, typical of the radiative zone of a B2 star (Brun, private communication; see also Augustson, Brun \& Toomre\,\citealp{augustson11}) and a degree of differential rotation ($ q\equiv r\vert\nabla\Omega\vert/\Omega$; Eq.~(13) of Spruit\,\citealp{spruit99}) equal to 1 following Spruit~\citet{spruit99}. We obtain that the initial critical field is $B_{\rm crit,init}$$\sim$430~G. Using again the ratio provided by Braithwaite\,\citet{braithwaite08}, but this time between the radiative zone and the surface, i.e. typically a factor 10, the critical initial surface field is $B_{\rm crit,init,surf}$$\sim$40~G. The corresponding current critical surface field is below this value, because of dissipation caused by the ohmic diffusion and the uniform rotation state caused by the Lorentz torque. Again, the Spruit criterion is based on approximations. Nevertheless, it shows that the surface magnetic field observed in V2052~Oph (400~G) is sufficient to explain the observed non-axisymmetric magnetic configuration and lack of mixing associated to uniform rotation.

More precise criteria extracted from detailed modelling, in particular taking into account the full coupled transport processes and the geometrical configuration (for example following Mathis \& Zahn\,\citealp{mathis05}; Spada, Lanzafame \& Lanza\,\citealp{spada10}; Duez \& Mathis\,\citealp{duez10a}; Duez, Braithwaite \& Mathis\,\citealp{duez10b}), are required to determine more precisely the level of magnetic field necessary to inhibit mixing in B stars. Nevertheless, our first estimate with both criteria above confirm that it is probably the presence of a fossil magnetic field in V2052~Oph that reduces the core overshooting parameter, i.e. the amount of non-standard mixing processes. 

As a consequence of our findings, the usual interpretation that rotational mixing is responsible for the N-enrichment observed in early B-type stars (Meynet et al.\,\citealp{meynet11}) cannot be applied to V2052~Oph. Preliminary results of Bourge et al.\,\citet{bourge07} showed that radiatively-driven microscopic diffusion may be an alternative explanation for this abundance peculiarity in V2052~Oph and other targets. Complete diffusion computations including radiative forces in $\beta$~Cep-type stars, which are not yet available, should include magnetic fields as there seems to be a higher incidence of nitrogen overabundance in stars with a detected magnetic field of the same order (i.e. a few hundred Gauss) as the one detected in V2052~Oph (Morel et al.\,\citealp{morel08}). 

\section{Conclusions}\label{sect_conclusions}

Our study was based on extensive ground-based, high-resolution, high-S/N, multisite spectroscopic measurements spread over many years. Similar efforts were already performed for two other $\beta$~Cep stars, $\nu$~Eri and 12~Lac, leading to around ten independent pulsation frequencies. For V2052~Oph, our campaign only led to the detection of three pulsation modes but, contrary to the two other cases, the rotation period is also found, which provides a strong additional constraint. The difficulty to detect more modes in V2052~Oph 
may be explained by lower amplitudes of the modes due to a higher rotation rate compared to the two other stars.

Comparing our identifications of the degrees of the modes with the ones coming from the photometry, we are confident that the dominant mode is radial while the two additional modes have $\ell$=$4$. Moreover, combining the outcome of the FPF method with that of our basic modelling, we conclude that the non-radial modes are prograde $(\ell,m)$=$(4,2)$ or $(4,3)$ components of two multiplets with $n$=$-2$ and $n$=$-3$. The surface equatorial rotational velocity was also deduced as $v_{\rm eq}$=73$\pm$2 km~s$^{-1}$.

The observation of three independent modes with constraints on their wavenumbers along with the rotation period was enough to deduce ranges for the mass ($M \in [8.2,9.6]\ $M$_\odot$), central hydrogen abundance ($X_c \in [0.25,0.32]$) and other parameters of V2052~Oph that are furthermore fully compatible with the spectroscopic $T_{\rm eff}$ and $\log g$, and with the inclination angle $i$ determined from the modelling of the magnetic field. More importantly, we showed that only models with no or mild core overshooting could satisfactorily reproduce our observational properties.


We made a comparison with $\theta$~Oph because it is an asteroseismically modelled $\beta$~Cep star having similar fundamental parameters as our object of interest. We would expect a larger core overshooting parameter value in the faster rotator but the contrary is found. It can be explained by the fact that V2052~Oph is a magnetic star while $\theta$~Oph is not. Indeed, using two approximate criteria, we showed that the magnetic field present in V2052~Oph is strong enough to inhibit non-standard mixing processes in its interior. In order to check if the internal rotation of V2052~Oph is indeed uniform or not, we would need additional constraints such as those obtained from a high-precision light curve as given by the CoRoT and Kepler satellites. Indeed, space-based photometry has already revealed many pulsation frequencies, even in objects that were thought to be monoperiodic from the ground (Aerts et al.\,\citealp{aerts06}; Degroote et al.\,\citealp{degroote09}).

\section*{Acknowledgments}
GH would like to thank Michael Endl for his assistance during the observations which made them successful. CN and SM thank Jean-Paul Zahn for fruitful discussions. The research leading to these results has received funding from the European Research Council under the European Community's Seventh Framework Programme (FP7/2007–2013)/ERC grant agreement n$^\circ$227224 (PROSPERITY). This work was also supported in part by PNPS (CNRS/INSU). TM acknowledges financial support from Belspo for contract PRODEX-GAIA DPAC. JE acknowledges finantial support from UNAM, PAPITT proyect IN122409. GH's contribution to this work has been supported by the Austrian Fonds zur F\"orderung der wissenschaftlichen Forschung under grant R12-N02. KU acknowledges financial support by the Spanish National Plan of R\&D for 2010, project AYA2010-17803.



\end{document}